# Direct phase mapping of the light scattered by single plasmonic nanoparticles

*Otto Hauler,[1] Frank Wackenhut,[1*] Lukas A. Jakob,[1] Alexander Stuhl,[1] Florian Laible,[2,3] Monika Fleischer,[2,3] Alfred J. Meixner,[1,3] and Kai Braun[1,3*]*

[1]Institute of Physical and Theoretical Chemistry, Eberhard Karls University, Tuebingen, 72076, Germany

[2]Institute for Applied Physics, Eberhard Karls University, Tuebingen, 72076, Germany

[3]Center for Light-Matter-Interaction, Sensors and Analytics LISA+, Eberhard Karls University, Tuebingen, 72076, Germany

In this work, we present a novel technique to directly measure the phase shift of the optical signal scattered by single plasmonic nanoparticles in a diffraction-limited laser focus. We accomplish this by equipping an inverted confocal microscope with a Michelson interferometer and scanning single nanoparticles through the focal volume while recording interferograms of the scattered and a reference wave for each pixel. For the experiments, lithographically prepared gold nanorods where used, since their plasmon resonances can be controlled via their aspect ratio. We have developed a theoretical model for image formation in confocal scattering microscopy for nanoparticles considerably smaller than the diffraction limited focus We show that the phase shift observed for particles with different longitudinal particle plasmon resonances can be well explained by the harmonic oscillator model. The direct measurement of the phase shift can further improve the understanding of the elastic scattering of individual gold nanoparticles with respect to their plasmonic properties.

I. INTRODUCTION

The interest in gold nanoparticles (GNPs) and their application is undiminished and has even further increased in recent years. They have been used for a broad range of applications, from the development of more efficient sensing, to LEDs [1], novel solar cells [2] or medical applications [3]. In biology, GNPs have been used to label individual proteins and their paths have been tracked through organisms [4]. In contrast to the more commonly used fluorescent dyes, GNPs have the advantage of being non-toxic and only slightly reactive. In addition, they do not bleach or blink. GNPs also have many applications in spectroscopic studies on single molecules, for example surface-enhanced Raman scattering [5]. With the plethora of possible applications, it is only natural that the methods for examining GNPs need to be developed to the same extent. Single particle sensing with GNPs is usually based on their luminescence emission [6], which intrinsically has a low quantum efficiency of $10^{-6}$ to $10^{-5}$ [7]. The scattering cross section of GNPs scales as the square of the particle volume and decreases drastically for smaller particles [8, 9]. To overcome these size restrictions detection schemes that scale linearly with the particle volume are desired. An example is the absorption cross section, which was exploited in photo thermal imaging [10]. Alternatively, the interference of the weak signal scattered by GNPs with a strong reference signal is also linear, but avoids excessive heating of the environment and allows the examination of particle properties. However, an important parameter is the phase of the scattered light and there are currently only few possibilities to experimentally study the phase shift of the light elastically scattered by nano objects, e.g. by using scanning near field microscopy [11, 12] or a Mirau interferometer [13]. Yet, the respective influence of the GNP properties on the strength of the phase shift is only little investigated and not yet well understood.

In this article, we introduce an extension of an inverted confocal microscope combined with a Michelson interferometer, enabling us to measure the relative phase shift $\delta$ caused by a single GNP in a diffraction limited focal volume. Furthermore, we will show that a damped harmonic oscillator model can adequately describe the relative phase shift $\delta$ with respect to the GNPs particle plasmon resonance. This new experimental method has the possibility to study numerous effects influencing the phase shift caused by a single GNP within the focal volume, which is fundamental to understand the elastic scattering properties.

## II. DESCRIPTION OF THE SETUP

In this work, we combine a confocal microscope with an additional beam path to implement a Michelson interferometer as illustrated in Figure 1.

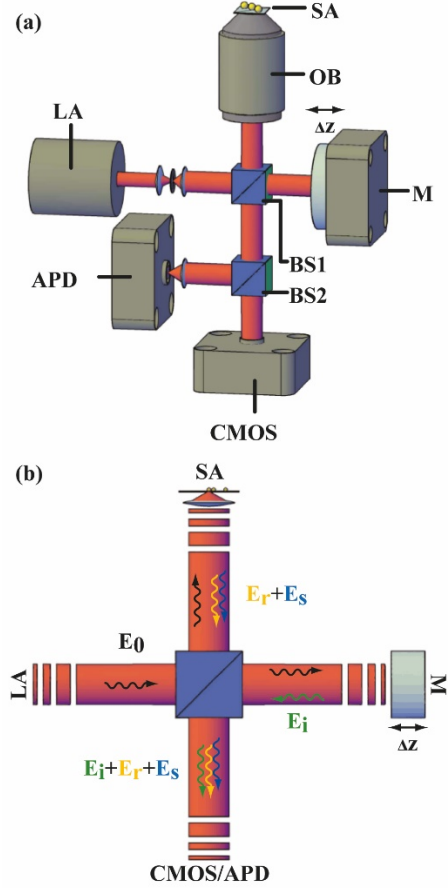

**Figure 1:** (a) Schematic drawing of our home-built confocal interference scattering microscope. LA: Narrow band diode laser with a wavelength of $\lambda = 632.8$ nm. BS1/2: Non-polarizing 50:50 beam splitters. OB: 1.25 NA oil-immersion objective. SA: Sample, consisting of a cover slip with single gold nanorods. M: Mirror mounted on a high precision piezoelectric actuator. APD: Avalanche photodiode. CMOS: Camera to control the superposition of the interfering beams. (b) shows the schematic beam path for confocal interference scattering microscopy. The excitation beam with an electric field component $\vec{E}_0$ is split at the beam splitter BS1 with half of its intensity focused onto the sample surface. The wave returning from the sample consists of two contributions, one that is reflected at the sample surface $\vec{E}_r$ and $\vec{E}_s$ which is elastically scattered by a single GNP. A moveable mirror M reflects the other half of the excitation beam with the electric

field component $\vec{E}_i$, forming the reference wave. The reflected, the scattered and the reflected reference waves are superimposed at the beam splitter BS1, and the optical path length of the reference beam can be varied by the mirror position $\Delta z$, leading to a path difference $\Delta L = 2\Delta z$.

Figure 1(a) shows a schematic drawing of our home-built inverted confocal sample scanning microscope, which is mounted on a damped optical table (RS4000 with IVP 2000 Newport) to reduce mechanical vibrations. Furthermore, a soundproof box around the microscope is used to minimize acoustic disturbances. The excitation source is a continuous wave laser diode with a wavelength of $\lambda_{ex} = 632.8$ nm and an external, grating stabilized, resonator (LA in Figure 1(a), DL100 with a SYS DC 110 control unit, Toptica Photonics). Its high coherence length of more than 300 m allows performing reliable interference measurements. A non-polarizing 50:50 beam splitter (BS1 in Figure 1(a)) is used to create the two arms of the Michelson interferometer, where the first half of the laser intensity is sent to the sample and the second half to a moveable mirror. The first interferometer arm consists of the objective lens (OB in Figure 1(a), CP-Achromat, Carl Zeiss MicroImaging GmbH, NA = 1.25), which is also used to collect the optical signal reflected or scattered by the sample (SA in Figure 1(a)). The reference beam is formed by the light passing straight through BS1 and reflected from the movable mirror (M in Figure 1(a)). The light intensity of the reference beam can be adjusted to match the intensity reflected from the sample surface by a neutral-density filter. The mirror M is mounted on a piezoelectric actuator (P-752.11C, Physik Instrumente GmbH (PI)), which allows to change the mirror position by $\Delta z$ in a controlled way with an extremely high repeatability of $\pm 1$ nm. Its linearity of 0.03% ensures a parallel orientation of the mirror's traverse path to the optical axis. As shown in Figure 1(b), the light reflected and scattered by the sample and the reference beam are merged at BS1. Altogether, the sample surface

SA, the moveable mirror M, and the beam splitter BS1 form a Michelson interferometer. The detection beam is split by a second beam splitter (BS2 in Figure 1(a)). The first beam section is focused on an avalanche photodiode (APD in Figure 1(a), SPCM-AQR by PerkinElmer Optoelectronics) to record scan images and interferograms while the second beam section is guided to a CMOS camera (CMOS in Figure 1(a), DCC1545 by Thorlabs) to ensure the overlap of the sample and reference beam. The sample is positioned reproducibly with a feedback controlled three-dimensional piezoelectric stage (P-517.3CL, PI)) having a scan range of $100 \times 100$ μm$^2$ and a positioning accuracy of one nanometer. A RHK SPM 1000 controller with the associated software XPMPro 2.0.0.0 (RHK Technology) is used to operate the whole setup. The software controls all components described above and links each scan point with a certain mirror position and the respective photon counts of the APD. The recorded datasets are visualized and analyzed in a self-developed toolbox written in MATLAB.

In this work, gold nanorods (GNRs) were investigated since their longitudinal particle plasmon mode can be easily manipulated by varying their aspect ratio (AR) [14–18]. The GNRs are prepared by electron beam lithography, as described in Gollmer et al. [19]. First, a 50 nm thick layer of indium tin oxide (ITO) is sputtered on a cleaned glass substrate. Afterwards, an approximately 110 nm thick film of polymethylmethacrylate (PMMA) is spin coated onto the sample and exposed to a focused electron beam with a scanning electron microscope (SEM, Philips XL30 by Philips GmbH with Xenos XPG2 pattern generator). The exposed PMMA is developed in a 1:3 mixture of methylisobutylketone and isopropanol. After removing possible organic residues with an oxygen plasma for 10 s, a 30 nm gold layer is thermally evaporated on the sample under high vacuum conditions. Finally, the remaining PMMA is removed by acetone, followed by a cleaning step in isopropanol and blow-drying with nitrogen.

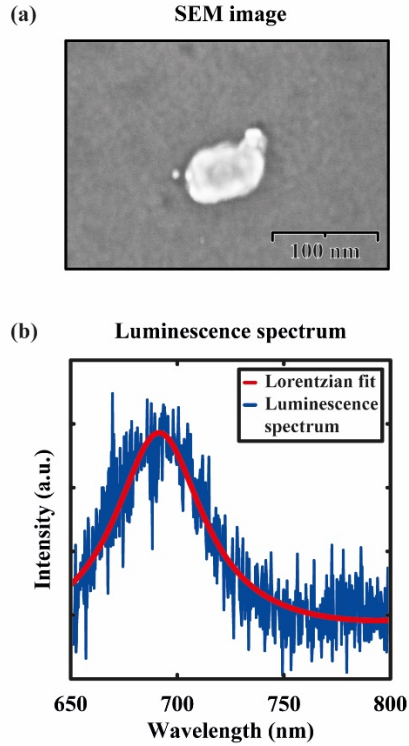

**Figure 2:** (a) SEM image of one GNR on ITO produced by electron beam lithography. The dimensions are 76 nm for the long axis and 50 nm for the short axis resulting in an AR of 1.5. (b) Photoluminescence spectrum of the GNR shown in Fig. 2 (a) with an intensity maximum at $\lambda_{max} =$ 687 nm and a full width at half maximum of 36 nm determined from the Lorentzian fit displayed in red.

Several arrays of 10 × 10 GNRs were fabricated. The long axis of the GNRs ranged from 70 nm to 140 nm, whereas the short axis was kept constant at 45 nm resulting in ARs varying from 1.5 to 3.1. Due to the resolution limits of electron beam lithography [20], deviations in length and width of ±5 nm are observed. The radius of curvature at the tips of the GNRs is approximately 15 nm. The GNRs were aligned parallel to each other with an intermediate distance of 5 μm. One exemplary GNR is shown in the electron micrograph in Figure 2(a) with its photoluminescence spectrum depicted in Figure 2(b). From the Lorentzian fit we determine an emission maximum of $\lambda_{em} = 687$ nm with a full width at half maximum of 36 nm for this particular GNR. Since the photoluminescence spectrum is determined by the particle plasmon, we can reasonably assume that its intensity maximum represents the spectral position of the longitudinal particle plasmon (PP) resonance of the GNR [21–23].

For small particles deposited on reflecting interfaces the signal collected by the objective lens from the diffraction limited excitation spot of the sample consists of two parts, the wave reflected at the sample surface (i.e. the respective glass-medium interface) $\vec{E}_r$, and the electric field component of the wave elastically scattered by the GNP $\vec{E}_s$, which is proportional to the incident radiation times the polarizability tensor [14, 24]. The intensity $I_{sca}$ is proportional to $|\vec{E}_r + \vec{E}_s|^2$, leading to three terms in equation (1), which are the square of the reflected field $|\vec{E}_r|^2$, i.e. the contribution of the constant background intensity, the square of the scattered field $|\vec{E}_s|^2$ and the interference term, which is proportional to the amplitudes of the reflected $|\vec{E}_r|$ and the scattered $|\vec{E}_s|$ field.

$$I_{sca} \propto |\vec{E}_r + \vec{E}_s|^2 = |\vec{E}_r|^2 + |\vec{E}_s|^2 + 2|\vec{E}_r||\vec{E}_s|\cos(\varphi) \tag{1}$$

The pure scattering intensity $|\vec{E}_s|^2$ scales with the square of the particle volume and is negligible compared to the other two terms. In conventional scattering microscopy, the image contrast is caused by the interference term in Equation (1) and sensitively depends on the phase shift φ between the reflected and the scattered light. This phase shift depends on the refractive index mismatch at the glass-air interface, the size and the shape of the scattering GNP, as well as on the dielectric medium surrounding the particle [15–18]. We propose an extended approach to the interferometric experiments based on the concept of Equation (1) by the use of a Michelson interferometer, which adds the reflected reference beam $\vec{E}_i$ as a third term to Equation (1). The phase shift between $\vec{E}_i$ and $\vec{E}_r$ can be varied by changing the position $\Delta z$ of the interferometer mirror M, which causes a path difference $\Delta L = 2\Delta z$ between the two interferometer arms. Analogous to the previous description of the image formation [14–18] in Equation (1), the intensity of the resulting signal $I_{sca}$ is now proportional to the absolute square of the sum of three electrical fields:

$$\begin{aligned} I_{det} &\propto \left|\vec{E}_r + \vec{E}_s + \vec{E}_i\right|^2 \\ &= \left|\vec{E}_r\right|^2 + \left|\vec{E}_s\right|^2 + \left|\vec{E}_i\right|^2 + 2\left|\vec{E}_r\right|\left|\vec{E}_s\right|\cos(\varphi) + 2\left|\vec{E}_r\right|\left|\vec{E}_i\right|\cos(k\Delta L) \\ &\quad + 2\left|\vec{E}_s\right|\left|\vec{E}_i\right|\cos(k\Delta L - \varphi) \end{aligned} \quad (2)$$

With the wave number $k = 2\pi/\lambda$ and the phase shift $\varphi$ of the reflected and scattered electric field. The last term in Equation (2) describes the interference of the wave scattered by the GNR $\vec{E}_s$ and reference wave and their respective phase difference $k\Delta L - \varphi$. Since $\Delta L$ can be controlled with high precision, $\varphi$ can be determined from $I_{det}$ recorded as a function of $\Delta L$. For illustration, we discuss first two special cases, i.e. when $\Delta L$ is set either to constructive ($\cos(k\Delta L) = 1$) or to destructive ($\cos(k\Delta L) = -1$) interference between the reflected and the reference wave. These two cases are illustrated in Figure (3), where we have acquired scan images for two fixed mirror positions.

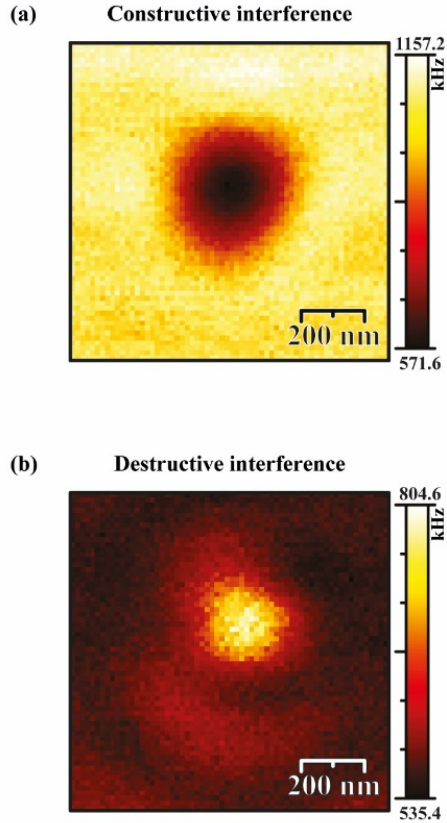

**Figure 3:** (a) and (b) depict confocal scattering images of the same GNP where the reference beam is set to constructive $(2\pi\Delta L/\lambda = n\pi$ with n = 0, 2, 4 ...) and destructive (n = 1, 3, 5 ...) interference, respectively. The intensities of the reflected and reference beam have been adjusted to have similar intensities, resulting in a seemingly conventional picture in (a), and a picture that is dominated by the scattering terms (b).

Figure 3 illustrates both cases, when the path difference $\Delta L$ is set to $n\lambda/2$ with either n = 0, 2, 4, ... or n = 1, 3, 5, ..., which gives a phase difference of even or odd multiples of $\pi$ between $\vec{E}_r$ and the reference beam $\vec{E}_i$, leading to constructive or destructive interference, respectively. The corresponding scan image in Figure 3(a) has a large background signal due to constructive interference of the wave reflected at the sample surface and the reference wave. The negative image contrast caused by the GNP directly shows that $\cos(\varphi)$ must be negative and hence $\varphi$ is between 90° and 270. However, fixing the path difference to $\Delta L = n\lambda/2$ with odd values of n results in destructive interference between the light reflected at the sample interface $\vec{E}_r$ and the reference beam $\vec{E}_i$. In this case, the background signal is reduced in the image shown in Figure 3(b), ideally leading to a background free image if the intensities of the reflected and reference beam are exactly the same. The intensity of the detected signal $I_{sca}$ is then only determined by the field component $\vec{E}_s$ scattered by the GNP, and depends on its optical properties. These results show that the detected intensity is clearly dependent on the position of the moveable mirror, which allows us to study the phase shift caused by an individual GNR by analyzing the intensity variation of the detected signal as a function of the path difference $\Delta L$.

For studying the phase shift as function of the particle plasmon resonance, we investigate individual GNRs with different ARs. In a first step we located the GNRs by conventional confocal scattering microscopy. A typical confocal scattering image of a single GNR with an AR of 1.5 (long axis 76 nm) at $\lambda_{ex} = 632.8$ nm with a scan area of $1 \times 1$ µm² and 64 sample points per line is shown in Figure 4 (a).

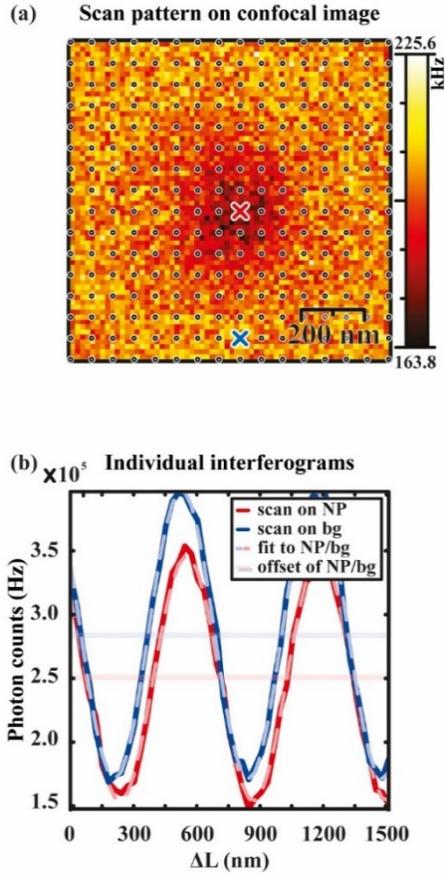

**Figure 4:** (a) shows a typical confocal scattering image at $\lambda_{ex} = 632.8$ nm of a single GNR with an AR of 1.5 (long axis 76 nm) with a scan area of $1 \times 1$ µm$^2$ and 64 sample points per line. The black points mark the positions, where interferograms were recorded. Fig. 4 (b) shows two interferograms (red and blue line) together with a sinusoidal fit (dashed grey lines); the first one is recorded with the GNR centered in the excitation as indicated by the red cross in Figure 2(a), and the second one is acquired with the focus on the glass coverslip (blue cross and line).

Afterwards, interferograms were recorded on a uniformly distributed grid of $16 \times 16$ pixels within the same scanning area (see Figure 4(a)). For each spatial position (black dots in Figure 4(a)), the moveable mirror M was scanned by a voltage ramp of 0 to 500 mV (corresponding to a movement of 750 nm) in intervals of 10 mV. Since the reference beam travels this distance twice the path difference ΔL is 1500 nm, resulting in 2.4 periods in the measured interferograms (Figure 4(b)) for a laser wavelength of 632.8 nm. The red and blue crosses mark the positions where the two interferograms displayed in Figure 4(b) were measured. The interferogram shown in red was acquired with a single GNR centered in the focus, and the blue one was acquired with the excitation focus on the glass coverslip. The difference in offset, amplitude, and phase between the interferograms shown in Figure 4(b) occurs since $\vec{E}_s = 0$ when the focal spot is centered on the blue cross and therefore the second, fourth and sixth term in Equation (2) are zero. Both interferograms are fitted with a sinusoidal wave function (grey dashed lines) to determine the background, the amplitude, the period and the phase of the interferogram. Rewriting Equation (2) we can express $I_{sca}$ as a function of the path difference:

$$I_{sca} \propto |\vec{E}_r + \vec{E}_s|^2 + |\vec{E}_i|^2 + 2|\vec{E}_i||\vec{E}_r + \vec{E}_s| \cos\left(k\Delta L - \tan^{-1}\left(\frac{|\vec{E}_s|\sin(\varphi)}{|\vec{E}_s|\cos(\varphi) + |\vec{E}_r|}\right)\right) \qquad (3)$$

Equation (3) shows that each interferogram can be fitted as a generic cosine function,

$$f(\Delta L) = y_0 + A \cos((2\pi/p)\Delta L + \delta). \qquad (4)$$

With $y_0$ as the offset, A as the amplitude, p as the period, and δ as the relative phase of the interferogram. The period p is given by the laser wavelength with $p = \lambda_{ex}$ (Figure 4 (b)) and is constant in all experiments. For data evaluation, all recorded interferograms underwent a computational procedure using a self-written Matlab algorithm for the sinusoidal curve fit.

III. RESULTS

Fitting Equation (4) to each interferogram of the 16 × 16 grid shown in Figure 4 (a) allows us to determine the offset $y_0$, the amplitude A, the period p and the phase δ as a function of ΔL for each pixel. The results are plotted in Figure 5, yielding images containing the respective interferogram information.

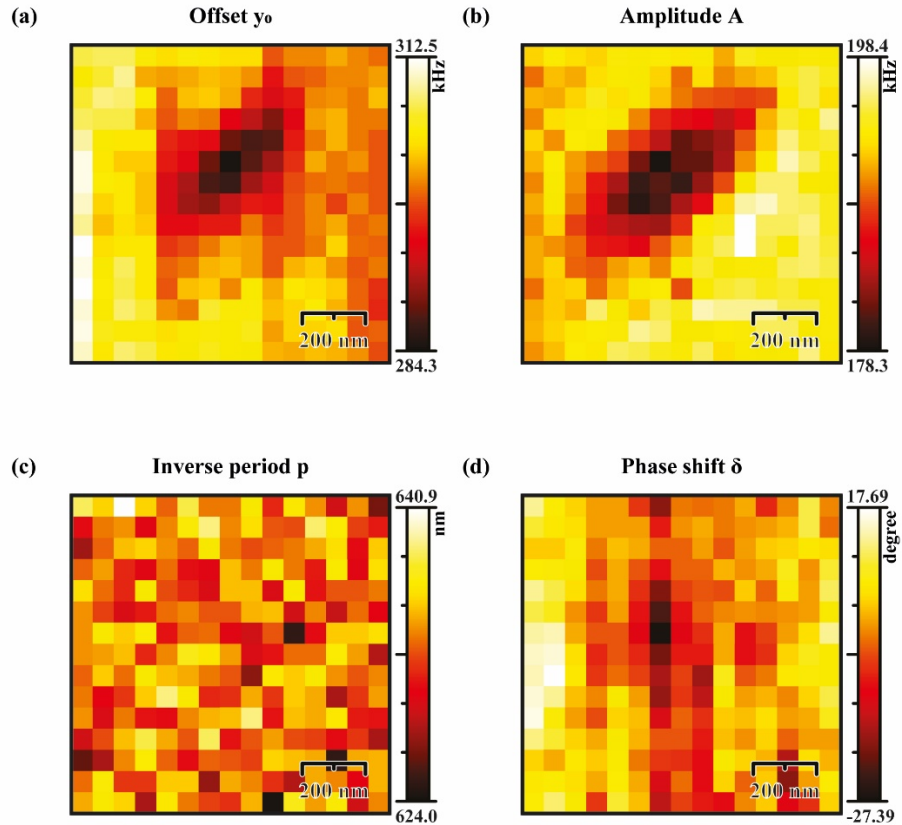

**Figure 5:** Illustration of the parameters obtained from a fit to the interferograms recorded at the positions marked by the black dots in Figure 4(a). The offset values $y_0$ and amplitudes A of the respective generic cosine functions are shown in (a) and (b), respectively. These two parameters result in images similar to the conventional confocal measurement in Figure 4(a). In Figure 5(c), the period p of the 16 × 16 interferograms is presented and gives a mean value of 633.52 nm in good agreement with the excitation laser wavelength. In (d), the results of the phase shifts δ for the fitted interferograms are shown. A 2D Gaussian fit yields a maximum relative phase shift on the GNR of $\delta = -15°$.

Figure 5(a) illustrates the offset $y_0$ and Figure 5(b) the amplitude A. These two parameters result in images similar to a conventional confocal scattering approach. An analysis of these images gives an image contrast of $-20$ kHz in Figure 5(a) and $-16$ kHz in Figure 5(b). This amounts to 6.5 % and 8.1 % of the background signal and is similar to the value of 11.6 % observed in the conventional scattering image. The period of the interferogram is presented in Figure 5(c), exhibits a mean value of 633.52 nm, and is equal to the excitation wavelength $\lambda_{ex} = 632.8$ nm within the error margin. This result confirms the system's high stability, as well as an overall reliable fit convergence. However, by our interferometric approach we also gain access to the phase shift δ of each interferogram, which is presented in Figure 5(d). In order to determine the phase shift δ, we fit a 2D Gaussian function to these images and the amplitude of this 2D Gaussian fit is the phase shift δ of the interferogram caused by a single GNR in the diffraction limited focal volume. For this particular example we observe a phase shift δ of -15°. However, it has to be noted that the observed phase shift δ is the phase shift of the interferogram caused by the GNR in the focal volume and not directly the phase $\varphi$ of the light scattered by the nanoparticle. δ is the phase shift as defined in Equation (4). The relation to the phase $\varphi$ is defined by comparing Equation (3) and Equation (4) and is

$$y_0 = \left|\vec{E}_r + \vec{E}_s\right|^2 + \left|\vec{E}_i\right|^2$$

$$A = 2\left|\vec{E}_i\right|\left|\vec{E}_r + \vec{E}_s\right| \tag{5}$$

$$\delta = -\tan^{-1}\left(\frac{\left|\vec{E}_s\right|\sin(\varphi)}{\left|\vec{E}_s\right|\cos(\varphi) + \left|\vec{E}_r\right|}\right)$$

In order to study the influence of the plasmon resonance on the relative phase shift δ we investigate individual GNRs with varying ARs and resonances from 650 nm to 900 nm determined as in Figure 2(b). The results are presented in Figure 6.

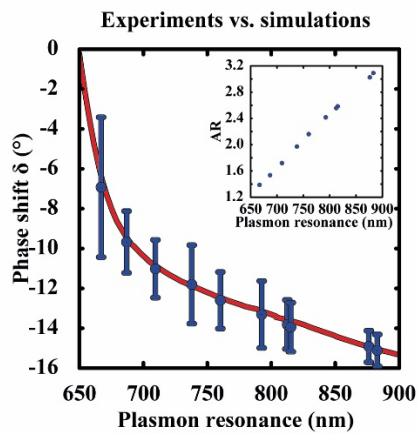

**Figure 6:** The blue dots represent experimental data for GNRs with different ARs and consequently different plasmon resonances. The error bars are the result of ten consecutive measurements and serve to confirm the reproducibility of our experimental set up. The red line illustrates the relative phase shift calculated with a damped harmonic oscillator approach. The observed phase shift δ becomes more negative for increasing wavelength differences between the excitation wavelength and the PP resonance. Inset: The experimentally recorded plasmon resonances are plotted against the ARs of the examined GNRs as blue dots. The ARs were obtained through SEM measurements of each individual GNR.

The blue dots in Figure 6 represent experimental data of the phase shift δ for GNRs as a function of the longitudinal PP resonance ranging from 667 nm to 883 nm, excited with a 632.8 nm laser. The longitudinal PP resonances were determined from the photoluminescence spectra of the respective GNR and the inset in Figure 6 shows the linear dependence of the PP resonance with the AR determined from SEM measurements. The error bars in Figure 6 represent the variation obtained from ten consecutive measurements of each GNR. The experimentally determined relative phase shift δ ranges from $-6.69°$ to $-14.87°$ and becomes more negative with increasing spectral difference between the excitation wavelength and the PP resonance. The red line displays a calculation based on an oscillator approach where the PP is treated as a damped harmonic oscillator periodically driven by an external periodical force. The equation of motion of such a driven oscillator can be described by an inhomogeneous differential equation:

$$\ddot{x} + 2\gamma\dot{x} + \omega_0^2 x = A_0 \cos(\omega t) \tag{6}$$

With x being the displacement, $\gamma$ the damping factor, and $\omega_0$ the resonance frequency of the oscillating system. $A_0$ is the amplitude and $\omega$ the angular frequency of the external periodical force. A solution to this differential equation for weak damping ($\gamma \ll \omega_0$) and stationary conditions can be written as:

$$x(t) = \frac{A_0}{\sqrt{(\omega_0^2 - \omega^2)^2 + 4\gamma^2\omega^2}} \cos(\omega t + \varphi) \tag{7}$$

The frequency dependent phase shift $\varphi$ between the excitation and the forced oscillation can be derived as:

$$\varphi = \tan^{-1}\left(\frac{\omega_0^2 - \omega^2}{2\gamma\omega}\right) - \frac{\pi}{2} \tag{8}$$

Accordingly, the phase shift $\varphi$ grows for $\omega < \omega_0$ from 0 to $-\pi/2$, is equal to $-\pi/2$ for $\omega = \omega_0$ and approaches $-\pi$ for $\omega \to \infty$. It is negative, i.e. the forced oscillation lags behind the excitation force. Equation (7) can be used to calculate the phase shift $\varphi$ of the harmonic oscillator relative to the excitation field. The resonance frequency $\omega_0$ of the scattering cross-sections and the damping factors $\gamma$ are calculated by using the boundary element method (BEM) [25] with specifications according to the geometry of each GNR determined by electron microscopy. The phase shifts $\varphi$ obtained by Equation (7), together with an additional phase shift of $\vec{E}_r$ caused by the total internal reflection at the sample surface due to the high NA of the objective lens [26], can be used to calculate the experimental phase shift $\delta$ on the basis of Equation (3). The red line in Figure 6 displays the result of these calculations and we find an excellent agreement with the experimental data, which leads to the conclusion that a simple harmonic oscillator model is suitable to describe the particle plasmon oscillation and its impact on the phase of the scattered light. Furthermore, these results show that this experimental approach allows to reliably study the influence of the AR of each GNR on the phase shift opening up a new detection scheme in single particle studies.

IV. CONCLUSION AND OUTLOOK

In summary, we have developed a new experimental method to observe the relative phase shift δ caused by a single GNP within a diffraction limited confocal volume. Furthermore, we present a theoretical model based on a driven harmonic oscillator to describe the observed experimental phase shifts. We have demonstrated that our approach is suitable to reliably determine the phase shift caused by single GNRs with different aspect ratios and systematically studied the influence of the relative spectral position of the PP resonance on the phase shift. This approach opens up a new and sensitive detection method in single particle sensing by exploiting elastic scattering. Single particle sensing usually uses the luminescence emission of GNPs, which intrinsically has a low quantum efficiency of $10^{-6}$ to $10^{-5}$ [7]. This drawback could be overcome by measuring the phase of the elastically scattered light. Moreover, the direct measurement of the phase shift improves the understanding of this not very well investigated aspect of elastic scattering of single GNPs.

V. AUTHOR INFORMATION

**Corresponding Author**

*Frank Wackenhut, frank.wackenhut@uni-tuebingen.de, Institute of Physical and Theoretical Chemistry, Eberhard Karls University, Auf der Morgenstelle 18, 72076 Tuebingen, Germany

*Kai Braun, kai.braun@uni-tuebingen.de, Institute of Physical and Theoretical Chemistry, Eberhard Karls University, Auf der Morgenstelle 18, 72076 Tuebingen, Germany

*Alfred J. Meixner, Alfred.meixner@uni-tuebingen.de, Institute of Physical and Theoretical Chemistry, Eberhard Karls University, Auf der Morgenstelle 18, 72076 Tuebingen, Germany

**Author Contributions**

The manuscript was written through contributions of all authors. All authors have given approval to the final version of the manuscript.

VI. ACKNOWLEDGMENT

The authors gratefully acknowledge funding by DFG grant BR 532/1-1, FL670/7-1 and ME 1600/13-3.

VII. ABBREVIATIONS

GNP, gold nanoparticle; GNR, gold nanorod; AR, aspect ratio; ITO, indium tin oxide; PMMA, polymethylmethacrylate; PP, particle plasmon.